# *What complexity science is, and why*

Petter Holme, Tokyo Institute of Technology

*This is an essay about understanding complexity science, via some peculiarities of the field, as a meeting place for a special kind of scientist. It comes out of my hobby of reading popular-science complex systems books, and builds on notes that have been collecting dust for almost a decade.*

**Definitions and disclaimers**

A striking feature of complexity science is the effort books and articles spend on defining the discipline itself, or complex systems (its study objects).[1] These definitions often come with disclaimers declaring themselves incomplete,[2] inconsistent with other definitions,[3] etc. In other words, claiming not to be the final word on the matter and preparing the readers for more definitions to come. But, the more definitions I read, the less well-founded the disclaimers seem. For the purpose of defining a scientific discipline, they are not only consistent,[4] but they also stand the test of time: "[B]y a complex system I mean one made up of a large number of parts that interact in a nonsimple way," wrote Herbert Simon in 1962. "[A] complex system is a collection of objects or agents that (a) have a high cardinality and (b) interact with one another in a non-trivial way," wrote Torres *& al.* almost 60 years later. All-in-all, I think the definers of complexity (in the broad, discipline-encompassing sense) actually agree with each other like one big happy family.

**Language and identity**

So there are two outstanding questions: (i) why so many authors feel compelled to state definitions of complex systems, and (ii) why these are accompanied by disclaimers about how ill-defined the field is.

I believe (ii) results from a sloppy use of "complexity"—the same word denoting the field itself and a multitude of more or less measurable, more or less system-specific phenomena.[5] We cannot define a field as a collection of discovered phenomena—a definition needs to be flexible and broad enough to allow the field to grow in yet unknown directions.[6] Neither can we assume there is a profound principle that manifests itself as all these phenomena and the field being the study of this principle—that would be a just too esoteric way of defining something as concrete as a discipline.[7]

Point (i) is to some extent explained by (ii), but there

---

1. Of course, almost every field has an ongoing discussion about what it is and should be—sometimes heated, sometimes prolonged for centuries. Still, when it comes to stating definitions of its study object, complexity science wins the gold medal.

2. "I shall not undertake a formal definition of 'complex systems'." (Simon 1962)

    "Nobody knows quite how to define it, or even where its boundaries lie" (Waldrop 1993)

3. "Distinct definitions exist, and not one is globally agreed upon" (Torres *& al.* 2021)

    "There is no agreement about what complexity is, whether it can be measured, and, if so, how, and there is no agreement about whether complex systems all have some common set of properties." (Ladyman, Wiesner 2020)

    "Complexity, like life and consciousness, does not have a rigorous definition." (Holland 2014)

4. Sometimes, the authors add requirements (unnecessary, I think[4a]) that the interactions should lead to something worthy of research, but otherwise, general definitions roughly follow that form[4b].

4a. Like emergence. For example, Torres *& al.* (2021) continue by requiring the system to show emergent properties. I think that is unnecessary because finding out whether a complex system shows emergence should also count as complexity science—we need a criterion based on what the system is, not how it behaves. Whether philosophy or science, the relationship between complexity and emergence is a fascinating topic, though (*cf.* Page 2011).

4b. In the 1980s and 90s literature, there is another seemingly different line of definitions (Pagels 1988, Lewin 1993)—that complex systems are what lies between order and chaos. At the highly coarse-grained level needed when defining a scientific discipline, I think these definitions are more or less the same. Note that "chaos" takes a much fluffier meaning (often, but not always, simply "randomness") here than deterministic chaos of dynamical systems. I also assume we have a sufficiently generous definition of "non-trivial interactions."

5. Lloyd (2001) lists over 40 measures capturing different aspects of complexity in various systems. Mitchell (2005) lists several manifestations of complexity—chaotic dynamics, statistical measures (descendants of Kolmogorov complexity), fractal structures, emergence, logical depth, and whatnot. As I see it, such measures and manifestations are research topics of complexity science but little related to the definition of a field. In a passage that I think shows this confusion Mitchell (2005) writes: "If the *sciences* of complexity are to become a *science* of complexity, then people are going to have to figure out how diverse notions of *complexity*—formal and informal—are related to one another, and how to most usefully refine the overly complex notion of complexity." The final italicized "complexity" refers, as I understand, to the definition of the field, and isn't "overly complex" unless one confuses it with the is also another reason. Complexity science belongs to a division of science perpendicular to the academic one.[8] It (presently) has no neighbors along this perpendicular axis, so we can't define complexity science by relating it to the existing disciplinary boundaries. Thus, it's challenging to maintain a disciplinary identity as a complexity scientist, so stating and discussing the definition serves to unite and tighten the community.[9]

### An unusual science for unusual scientists

The above discussion raises yet another question: How can the current situation be stable? Why don't all scientists with a background in traditional disciplines simply take the results they need back to their home fields and give up on calling it complexity science? Or why don't people call themselves specialized complexity scientists and close the door to interdisciplinary collaborations?[10] Once again, I would have preferred not to rely on a behavioral or social explanation,[11] but it simplifies the situation to such a degree that we don't even have to summon Occam: *Complexity science is driven by researchers loving science in all its hues—scientists who want to contribute to, be influenced by, and collaborate across more than just one disciplinary boundary.*

This proposition primarily comes from my own experience, field observations, and some very readable books—in particular Mitchell Waldrop's (1993) and Roger Lewin's (1992) books about the early days of the Santa Fe Institute (both called *Complexity*, but with different subtitles). They both tell the story as a torch relay through the sciences run by the legends of complex systems. Brian Arthur meets Stuart Kauffman and gets an epiphany, Stuart Kauffman meets John Holland and gets an epiphany, and so on. Other books on the same topics, George Cowan's memoirs (2010), Kauffman's *At Home in the Universe* (1993), and Gell-Mann's *The Quark and the Jaguar* (1994), all paint a similar picture. That of celebrated scientists (though something of mavericks in their home fields) getting their best and most rewarding ideas when discussing science across disciplinary borders. In other words, complexity science is first and foremost a meeting place for pandisciplinary-minded[12] scientists.

### The rest follows

Given a description of complexity science as a pandisciplinary discussion forum, many idiosyncrasies follow

"diverse notions of complexity" (of particular systems). Moreover, I don't see the point of the "sciences" becoming "a science" since the rest of Mitchell's (excellent) book discusses complexity science as one. In general, complexity science usually presents itself as a unit and functions thus in the larger academic context. Hence, I interpret sciences (plural) as the intersection of complexity science and the traditional disciplines, and with those adjustments, the passage makes sense to me.

6. Several authors (Byrne & Callaghan 2014, Ch. 2) pointed out that if one only defines complexity as what can emerge from simple models or rules, some areas inaccessible to traditional, reductionist science will be out of reach also for complexity science. At the same time, complexity science presents itself as bold and visionary, so of course, definitions are not meant to constrain the field unnecessarily.

7. Searching for such principles and laws could, of course, be a research topic within the field, *e.g.* Kauffman (1993) argues that it is the main challenge.

8. My working title for this text was "Parallel lives meet at complexity," a play on "parallel lines [boundaries between academic disciplines] meet at infinity [*i.e.*, never]." Also, capturing the interdisciplinary joy of meeting people with similar experiences from other fields. (The current title is tongue-in-cheek and a bit of an hommage to the audacity of *How Nature Works*, Bak 1996.)

9. It is tempting to make religious analogies, like pointing out that ruminatively asking "what is complexity?" could lead to enlightenment, or that the definition of the field functions as a creed. But I think we better stay away from a spiritual language not to give the impression that complexity science wants to be judged by other standards than good old empirical science.

10. Complexity science has designated institutes and university centers, journals, conferences, and societies. AFAIK, they all emphasize its interdisciplinary nature, to the extent that the graduates from complexity science programs primarily call themselves interdisciplinary rather than specialized in complexity science. (There are probably exceptions here and hard to be very categorical.)

11. It feels unsatisfactory to rely on the scientists themselves and their motivations rather than deducing all descriptions from scientific precepts. We want complexity science to be defined by the outer reality rather than its members. It is still very much true—like many authors pointed out (*e.g.*, Solé & Goodwin 1988, Gell-Mann 1994)—that reductionist science will inevitably leave open questions behind for a discipline like complexity science. But what people study, who they are, and how they present their research is much easier to explain as socio-psychological phenomena.

Relatedly, other fields could, of course, be described by their social organization and maybe also the personality of its members. However, for complexity science, I think such a description explains, relatively speaking, more.

Finally, I use "personality" in a very sweeping sense, including scientific ambitions and interests. There is no definite causal direction—people's personalities affect their choice of research direction, but the science one gets involved in influences oneself as well.

12. Maybe "multidisciplinary" is better, but then with "multi" meaning "more than just two."

without too many leaps of faith. Since reductionism focuses on describing one unit and its interaction with its environment, a "large number of parts that interact in a nonsimple way" would feel unconventional in many branches of mainstream science. Still, it is close enough to home to come across as meaningful to mainstreamers. The intersection of topics interesting to all traditional disciplines might not be too far from complexity science.[13]

An unorthodox scientific discipline is itself a complex system, but of that other kind—where, yes, there are emergent phenomena, but no concise description (Byrne & Callaghan 2014, Ch. 2). This essay gives my mental model of the field but is far from complete.[14]

Finally, I wish this text would naturally build up to a roadmap for the future. It doesn't, but I can make some fluffy forward-looking notes. One message would, of course, be that it's about time to put the discussion of definitions aside and do science (a reminder to myself to stop writing texts like this). Another to stop looking into the past and see complexity science for what it could be and not what has been. Finally, to keep the sense of interdisciplinary camaraderie from the foundational years. There will always be people with a vast intellectual scope among whom the chance of finding creative and original thinkers is exceptionally high. Complexity science's best move so far was its first—to present itself as a haven for these people.

13. Here it would be easy to derail into a discussion about universality (Buchanan 2000)—the idea, based on an analogy from statistical physics, the same fundamental mechanism could explain very different phenomena throughout the sciences. The most well-known such mechanism is self-organized criticality (Bak 1996). This connection is material for a npaper or book. My standpoint is that universality is important, but so are non-universal, system-specific phenomena. Thus, we should not let the search for it be a research direction *per se*, only report it as we come across it.

14. One additional issue is that complexity science is fundamentally computational (Pagels 1988). Even if we build analytical theories for much of complexity science (Boccara 2003), I think computational thinking takes a more central position, and I'm not only referring to the most-is-actually-computation undercurrent (Flake 1998).

    Another notable feature is the many forms and levels of modeling complexity science employs, which set it aside from just being a statistics of emergent phenomena (Holland 2012).

    A third point worth covering is how to relate complexity science to systems science—the unruly aunt that favors intricate systems diagrams over minimal toy models, feedbacks over universality, engineering over science. In some of my blog posts, I tried to capture the continuum between these intellectual traditions. My hunch (and mini-survey) says that most complexity scientists treat them as one. Still, complexity science textbooks rarely mention systems science, let alone describe it (among the references, Sayama 2015, is an exception). Moreover, complexity-science books laud minimal models and complexity emerging from simple models to the extent that they rule out much of systems science.